\documentclass[11pt,twoside,english]{article}
\usepackage[T1]{fontenc}
\usepackage{lmodern}
\usepackage[cp1252,utf8]{inputenc}
\usepackage{geometry}
\usepackage[english]{babel}
\usepackage{textcomp}
\usepackage{amsmath}
\usepackage{amssymb}
\usepackage{cancel}
\usepackage{graphicx}
\usepackage{microtype}
\usepackage{color}
\usepackage{fancyhdr}
\usepackage[unicode=true,
 bookmarks=true,bookmarksnumbered=false,bookmarksopen=false,
 breaklinks=false,pdfborder={0 0 1},backref=false,colorlinks=false]
 {hyperref}
\hypersetup{pdftitle={Example-driven Introduction to the Differential Algebra Normal Form Algorithm using the Centrifugal Governor},
 pdfauthor={Adrian Weisskopf},
 pdfsubject={Introduction DA Normal Form Algorithm},
 pdfkeywords={Differential Algebra, Normal Form Algorithm, Centrifugal Governor}}

\begin{document}

\title{Introduction to the Differential Algebra Normal Form Algorithm using the Centrifugal Governor as an Example}

\author{Adrian Weisskopf\footnote{Michigan State University - Department of Physics and Astronomy | weisskop@msu.edu}}
\maketitle
\begin{center}
{\Large MSUHEP Report: 190617}
\end{center}
\begin{abstract}
This paper provides a detailed introduction into the differential algebra (DA) based normal form algorithm using the example of the symplectic one dimensional system of the centrifugal governor. The intention of this paper is to make the single steps of the algorithm as clear as possible in the hope that the understanding and use of DA normal form methods are spread throughout the scientific community.
\end{abstract}

\tableofcontents

\newpage

\pagestyle{fancy}
\fancyhead[RE,LO]{Introduction to the DA Normal Form Algorithm}
\fancyhead[LE,RO]{Weisskopf}
\fancyfoot[C] {\thepage}
\fancyfoot[LE,RO]{\href{https://bt.pa.msu.edu/cgi-bin/display.pl?name=NFCGExample}{MSUHEP Report: 190617}}
\setlength{\headheight}{13.6pt} 

\section{Introduction}

This paper provides an example-driven walk-through of the differential algebra (DA) based normal form algorithm for a symplectic one dimensional (1D) system. 

The DA framework was first developed to its current extent by Berz \textit{et al.} \cite{Berz1999MMM,Berz1987PowerSeriesTracking,Berz1988DABeam}. It allows for an analytic/numeric-hybrid representation and manipulation of analytic functions in form of polynomial expansions. Taylor polynomials of order $m$ are represented by DA vectors which store the coefficients of the individual terms with floating point accuracy together with the order of the corresponding variables. Operations are defined on the DA vectors and implemented in the framework. 

\begin{figure}[!ht]
\vspace{0pt}
\centering{}%
\begin{tabular}{c|c}
\includegraphics[width=0.56\textwidth]{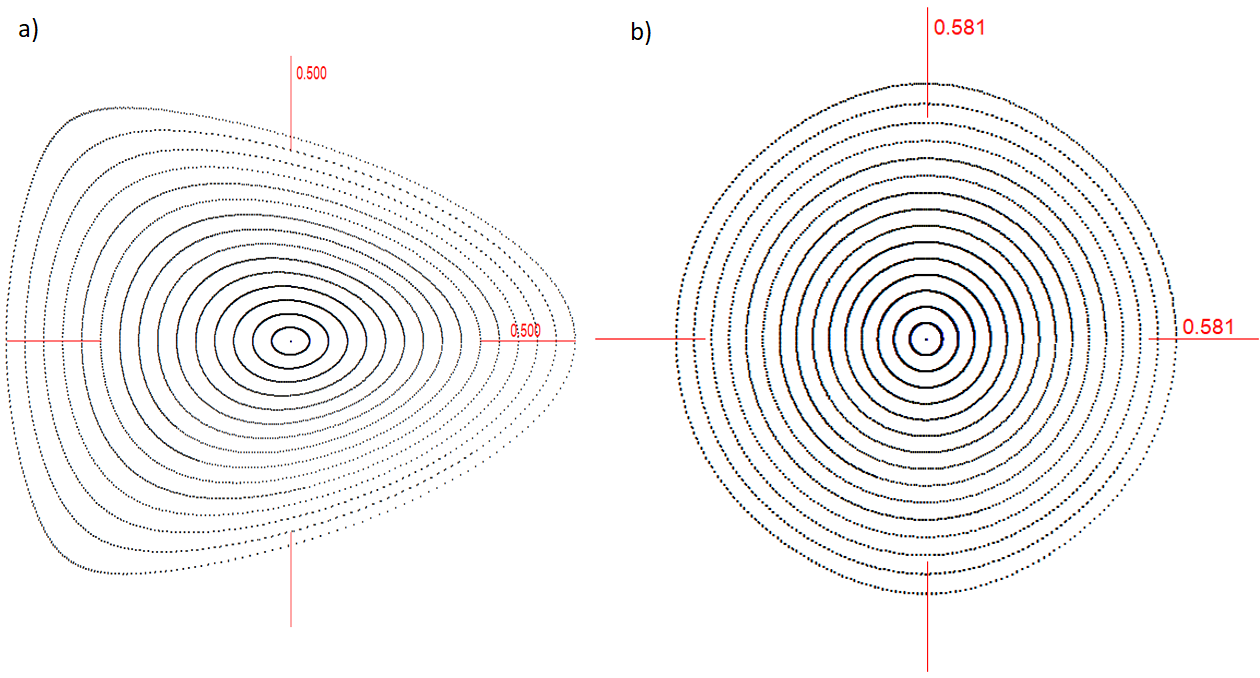}\includegraphics[width=0.05\textwidth]{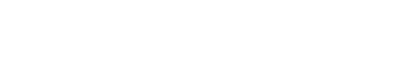} & \includegraphics[width=0.05\textwidth]{White_space.PNG}\includegraphics[width=0.28\textwidth]{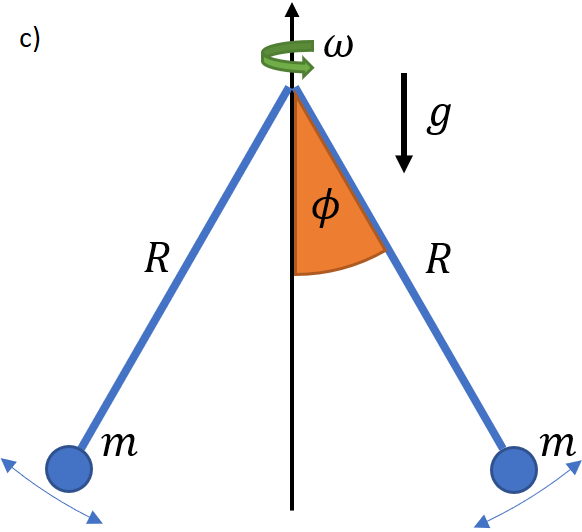}
\end{tabular}
\vspace{0pt}
\caption{a) Phase space tracking of Poincaré map. b) Phase space tracking of associated normal form map shows circular behavior. c) Illustration of centrifugal governor.}
\label{fig:Tracking+NF}
\end{figure}

The DA normal form algorithm \cite{Berz1999MMM} is an advancement from the DA-Lie based version, the first arbitrary order algorithm by Forest, Berz, and Irwin \cite{forest1989normal}. The DA normal form algorithm yields a nonlinear change of phase space variables by an order-by-order transformation that significantly simplifies the representation to be rotational invariant up to the order of calculation. Thus, the resulting normal form representation, up to the order of calculation, constitutes circular motion with quasi-invariants as radii and only normal form phase space amplitude (and parameter) dependent angle advancements (see Fig. \ref{fig:Tracking+NF}a+b).
The explanations on the DA normal form algorithm in this paper are based on Ref. \cite{Berz1999MMM}, which also contains details on the calculation of the DA normal form algorithm for higher dimensional systems and other various special cases like non-symplectic systems. The appendix \ref{sec:Appendix} yields a short summary on the main changes when considering higher dimensional phase space systems, e.g. resonances etc.

We start with an introduction to the centrifugal governor and the derivation of the equations of motion. This is followed by the calculation of an approximation of the flow of the equations of motions around an equilibrium position/point. This approximated flow is referred to as fixed point map. Given a linearly stable fixed point map, we go through the DA normal form algorithm step by step to obtain the rotationally invariant normal form map. The normal form is calculated up to third order. It will become obvious during the example that transformations of higher even and odd orders follow the same procedure as the second and third order transformation, respectively.

\section{The centrifugal governor}

The centrifugal governor is a device involving gravitational and centrifugal forces with the rotation axis parallel to the direction of the gravitational force. We are considering a mathematically idealized governor, which consists of two massless rods of equal length $R$ suspended at opposite sides of the rotation axis and a point mass $m$ attached at the end (opposite to where the rod is mounted) of each of the rods (see Fig. \ref{fig:Tracking+NF}c). The angle between the rotation axis (in direction of gravitational pull) and the rod is denoted with the angle $\phi$. 

At a sufficiently high rotation frequency $\omega>\omega_{min}$ (see eq. \ref{eq:equipoint}) the centrifugal force $F_{cent}$ carries the masses radially outward, while the gravitational force $F_{grav}$ pulls down on them. The tension $T$ in the rods is given by the vector sum of the two forces. For every rotation frequency $\omega>\omega_{min}$, there is an equilibrium angle $\phi_{0}>0$ (see subsection \ref{subsec:Equilibrium-point}) where the gravitational and centrifugal forces balance. 

A centrifugal governor is usually used in the application of a self-regulating mechanism, where the angle $\phi$ is negatively coupled back to the rotation frequency $\omega$ through an extra mechanism (increase in $\phi$ $\Rightarrow$ decrease in $\omega$). Accordingly, in those applications, e.g. the steam engine, the rotation frequency $\omega$ changes during the regulating process.

For the introduction to the normal form algorithm, we are considering the motion of the system with a \textbf{fixed} externally driven rotation frequency $\omega$. In particular, we are interested in the motion around the equilibrium position/angle $\phi_{0}\left(\omega\right)$ for a specific rotation frequency $\omega$. If the rods are perturbed from their equilibrium position/angle $\phi_{0}\left(\omega\right)$, they oscillate around $\phi_{0}\left(\omega\right)$ and we what to study the dynamics of that motion using the DA normal form algorithm.

\subsection{Equilibrium point\label{subsec:Equilibrium-point}}

At the equilibrium point/angle $\phi_{0}>0$ of the centrifugal governor for $\omega>\omega_{min}$, the vertical forces $mg=T\cos\phi_{0}$ and radial forces $m\omega^{2}R\sin\phi_{0}=T\sin\phi_{0}$ balance out in each of the rods. Rewriting the tension $T$ in terms of the radial force equilibrium $\left(T=m\omega^{2}R\right)$ and using this expression in the vertical equilibrium yields: 
\begin{align}
mg & =m\omega^{2}R\cos\phi_{0}\nonumber\\
\cos\phi_{0} & =\frac{g}{\omega^{2}R}\nonumber\\
\phi_{0} & =\arccos\left(\frac{g}{\omega^{2}R}\right) \quad \text{for} \quad \omega>\sqrt{\frac{g}{R}}=\omega_{min}.\label{eq:equipoint}
\end{align}
We are using $k$ to summarize the parameters $g$, $R$ and $\omega$ with
\begin{equation}
	k=\frac{\omega^2R}{g}\quad\text{and}\quad k_{min}=\frac{\omega_{min}^2R}{g}=1\label{eq:gk=00003Dw2R}.
\end{equation}
Cases with $\omega\le\omega_{min}$ or $k\le1$ are not considered, since the gravitational pull dominates over the centrifugal forces for those cases such that the masses are just be hanging straight down $\left(\phi_{0}=0\right)$. 

The following list yields a couple of equilibrium angles $\phi_{0}$ for various $k>1$:
\begin{align}
\phi_{0}\left(k=\frac{2}{\sqrt{3}}\right) & =30\text{\textdegree}=\frac{\pi}{6}\nonumber\\
\phi_{0}\left(k=\sqrt{2}\right) & =45\text{\textdegree}=\frac{\pi}{4}\nonumber\\
\phi_{0}\left(k=2\right) & =60\text{\textdegree}=\frac{\pi}{3}\nonumber\\
\phi_{0}\left(k=\frac{1}{2}+\frac{\sqrt{5}}{2}\right) & =51.273\text{\textdegree}\nonumber\\
\lim_{k\rightarrow\infty}\phi_{0}\left(k\right) & =90\text{\textdegree}\nonumber.
\end{align}

\subsection{Equations of motion}

The Lagrangian formulation of the problem yields
\begin{align}
l=\frac{L}{m} & =\frac{\dot{\phi}^{2}R^{2}+\omega^{2}R^{2}\sin^{2}\phi}{2}-gR\left(1-\cos\phi\right),\nonumber
\end{align}
where $U=gR\left(1-\cos\phi\right)$ is the gravitational potential and $U_{eff}=U-\frac{\omega^{2}R^{2}\sin^{2}\phi}{2}$ is the effective or centrifugal-gravitational potential. 

The generalized canonical momentum $p_{\phi}$ to the position variable $\phi$ is given by
\begin{equation}
p_{\phi}=\frac{dl}{d\dot{\phi}}=R^{2}\dot{\phi}.\nonumber
\end{equation}

Using the Legendre transformation, the Hamiltonian and the equations of motion (Hamilton equations) are obtained:
\begin{align}
H & =\frac{p_{\phi}^{2}}{2R^{2}}+gR\left(1-\cos\phi\right)-\frac{\omega^{2}R^{2}\sin^{2}\phi}{2}\nonumber\\
\dot{\phi}=\frac{dH}{dp_{\phi}} & =\frac{p_{\phi}}{R^{2}}\label{eq:eomq}\\
\dot{p}_{\phi}=-\frac{dH}{d\phi} & =-gR\sin\phi+\omega^{2}R^{2}\sin\phi\cos\phi\nonumber\\
 & =gR\sin\phi\left(k\cos\phi-1\right),\label{eq:eomp}
\end{align}
where the last step used the relation $\omega^{2}R=kg$ from eq. \ref{eq:gk=00003Dw2R}.

\subsubsection{Units}

Distances are considered in units of the rod length $R[m]$ such that $R[R]=1$ and time is considered in units of $T_{0}[s]=\sqrt{\frac{R[m]}{g[m/s^{2}]}}$ such that the gravitational constant $g[R/T_{0}^{2}]=1$. Accordingly, $\omega$ is considered in units of $\Omega_{0}[1/s]=\frac{1}{T_{0}[s]}=\sqrt{\frac{g[m/s^{2}]}{R[m]}}$, with $\omega_{min}[\Omega_{0}]=1$, and $k$ is considered in units of $K_{0}[1/s^{2}]=\Omega_{0}^{2}=\frac{1}{T_{0}^{2}}=\frac{g[m/s^{2}]}{R[m]}$, with $\omega^{2}[\Omega_{0}^{2}]=k[K_{0}]$. The canonical momentum $p_{\phi}$ is considered in units of $P_{\phi}[m^{2}/s]=\frac{R^{2}[m^{2}]}{T_{0}[s]}=\sqrt{R^{3}[m^{3}]\cdot g[m/s^{2}]}$. These dimensionless units are introduced to minimize the number of parameters in the system. Assuming $g[m/s^{2}]=9.81$ and $R[m]=0.0981$ yields $T_{0}[s]=0.1$, $\Omega_{0}[1/s]=10$, $K_{0}[1/s^{2}]=100,$ $P_{\phi}[m^{2}/s]=0.0962361$ and $\omega_{min}[\Omega_{0}]=1=10\,rad/s$, which corresponds to a bit more than 1.5 revolutions per second. 

\subsubsection{Expansion up to third order around equilibrium point}

As mentioned above, we are interested in the motion around the phase space equilibrium point, which is why we expand the equations of motion (eq. \ref{eq:eomq} and \ref{eq:eomp}) in the phase space variables $(\delta_{\phi},p_{\phi})$ around the equilibrium point $(\phi_{0},0)$. We are using the dimensionless units defined above with $R=1$, $g=1$ and $\omega^{2}=k$ and calculate the expansion up to third order, which is sufficient for the full 1D symplectic introduction to the DA normal form algorithms while keeping the number of terms to a minimum. The expansion of eq. \ref{eq:eomq} and \ref{eq:eomp} around the phase space equilibrium point up to third order yields
\begin{align}
\dot{\phi}=\dot{\delta}_{\phi} & =p_{\phi}\\
\dot{p}_{\phi} & =\sin\left(\phi_{0}+\delta_{\phi}\right)\cdot\left(k\cos\left(\phi_{0}+\delta_{\phi}\right)-1\right)\nonumber\\
 & =\sin\phi_{0}\left(k\cos\phi_{0}-1\right)+\delta_{\phi}\left(k\cos2\phi_{0}-\cos\phi_{0}\right)\nonumber\\
 & +\frac{\delta_{\phi}^{2}}{2}\sin\phi_{0}\left(1-4k\cos\phi_{0}\right)+\frac{\delta_{\phi}^{3}}{6}\left(\cos\phi_{0}-4k\cos2\phi_{0}\right)\nonumber\\
 & =0+\delta_{\phi}\left(\frac{1}{k}-k\right)-\frac{3\delta_{\phi}^{2}}{2k}\sqrt{k^{2}-1}+\frac{\delta_{\phi}^{3}}{6}\left(4k-\frac{7}{k}\right).
\end{align}

Writing the equations of motion in a vector notation and evaluating them for the case $\phi_{0}\left(k=2\right)=60\text{\textdegree}=\frac{\pi}{3}$ yields
\begin{align}
\begin{pmatrix}\dot{\delta}_{\phi}\\
\dot{p}_{\phi}
\end{pmatrix}=\vec{f} & =\begin{pmatrix}0 & 1\\
\frac{1}{k}-k & 0
\end{pmatrix}\begin{pmatrix}\delta_{\phi}\\
p_{\phi}
\end{pmatrix}+\begin{pmatrix}0\\
-\frac{3\sqrt{k^{2}-1}}{2k}\delta_{\phi}^{2}+\left(\frac{2k}{3}-\frac{7}{6k}\right)\delta_{\phi}^{3}
\end{pmatrix}\quad\text{for}\quad k>1,\nonumber\\
 & =\begin{pmatrix}0 & 1\\
-\frac{3}{2} & 0
\end{pmatrix}\begin{pmatrix}\delta_{\phi}\\
p_{\phi}
\end{pmatrix}+\begin{pmatrix}0\\
-\frac{3\sqrt{3}}{4}\delta_{\phi}^{2}+\frac{3}{4}\delta_{\phi}^{3}
\end{pmatrix}\quad\text{for}\quad k=2.\label{eq:eomk2}
\end{align}

\section{Map calculation via integration}

The following investigation of the system using the DA normal form algorithm is considering the centrifugal governor for $k=2$, given the equations of motions in eq. \ref{eq:eomk2}. The phase space variables of the fixed point map (the original variables) are denoted with $\left(q_{0},p_{0}\right)$ instead of $\left(\delta_{\phi},p_{\phi}\right)$ to limit subscripts to cases where they are necessary. The map is calculated by integrating the system of ODE's (eq. \ref{eq:eomk2}) from the initial phase space state 
\begin{equation}
\left(q_{ini},p_{ini}\right)=\left(\phi_{0}\left(k=2\right)+\delta_{\phi},p_{\phi}\right)=\left(\frac{\pi}{3}+q_{0},p_{0}\right)\nonumber
\end{equation}
until $t=1$. Note that, since any flow of the ODE's in eq. \ref{eq:eomk2} is always a fixed point map, the time of the integration can be chosen freely. The resulting map of the integration $\mathcal{M}_{0}=\left(Q\left(q_{0},p_{0}\right),P\left(q_{0},p_{0}\right)\right)^{T}$ has the following form: $\mathcal{M}_{0}=\mathcal{C}+\mathcal{L}+\sum_{m}\mathcal{U}_{m}$, where the constant part is denoted with $\mathcal{C}$, the linear part with $\mathcal{L}$ and each of the nonlinear parts of order $m$ with $\mathcal{U}_{m}$. Since the system is expanded around the equilibrium point $\phi_{0}\left(k=2\right)=\frac{\pi}{3}$, the constant part of the map has not changed from the starting equilibrium point: $\left(q_{const}=\phi_{0}\right)$. The following explicit formulation of $\mathcal{M}_{0}$ up to order three introduces the notation of various coefficients of the map:
\begin{eqnarray}
\mathcal{M}_{0}\left(q_{0},p_{0}\right) & = & \begin{pmatrix}\mathcal{M}_{0}^{+}\left(q_{0},p_{0}\right)\\
\mathcal{M}_{0}^{-}\left(q_{0},p_{0}\right)
\end{pmatrix}=\begin{pmatrix}Q\left(q_{0},p_{0}\right)\\
P\left(q_{0},p_{0}\right)
\end{pmatrix}\nonumber\\
 & = & \underbrace{\begin{pmatrix}q_{const}\\
p_{const}
\end{pmatrix}}_{\mathcal{C}}+\underbrace{\begin{pmatrix}\left(Q|q_{0}\right) & \left(Q|p_{0}\right)\\
\left(P|q_{0}\right) & \left(P|p_{0}\right)
\end{pmatrix}\begin{pmatrix}q_{0}\\
p_{0}
\end{pmatrix}}_{\mathcal{L}}\nonumber\\
 & + & \underbrace{\begin{pmatrix}\mathcal{U}_{2(2,0)}^{+}\\
\mathcal{U}_{2(2,0)}^{-}
\end{pmatrix}q_{0}^{2}+\begin{pmatrix}\mathcal{U}_{2(1,1)}^{+}\\
\mathcal{U}_{2(1,1)}^{-}
\end{pmatrix}q_{0}p_{0}+\begin{pmatrix}\mathcal{U}_{2(0,2)}^{+}\\
\mathcal{U}_{2(\text{0},2)}^{-}
\end{pmatrix}p_{0}^{2}}_{\mathcal{U}_{2}}\nonumber\\
 & + & \underbrace{\begin{pmatrix}\mathcal{U}_{3(3,0)}^{+}\\
\mathcal{U}_{3(3,0)}^{-}
\end{pmatrix}q_{0}^{3}+\begin{pmatrix}\mathcal{U}_{3(2,1)}^{+}\\
\mathcal{U}_{3(2,1)}^{-}
\end{pmatrix}q_{0}^{2}p_{0}+\begin{pmatrix}\mathcal{U}_{3(1,2)}^{+}\\
\mathcal{U}_{3(1,2)}^{-}
\end{pmatrix}q_{0}p_{0}^2+\begin{pmatrix}\mathcal{U}_{3(0,3)}^{+}\\
\mathcal{U}_{3(0,3)}^{-}
\end{pmatrix}p_{0}^3}_{\mathcal{U}_{3}}\label{eq:M11}
\end{eqnarray}

The position $Q$ and momentum $P$ components of the map $\mathcal{M}_0$ correspond to the upper and lower component and are denoted with \lq +\rq\,and \lq -\rq\,, respectively. The coefficients in the upper and lower component for the nonlinear $k(=a+b)$th order terms $q^{a}p^{b}$ are denoted with $\mathcal{U}_{k(a,b)}^{\pm}$. The coefficients in the linear matrix $(a|b)$ indicate the factor with which $a$ is linearly dependent on $b$. 

The following table \ref{tab:M0} lists the values of the coefficients in equation \ref{eq:M11} above for the centrifugal governor example of $k=2$. \\
\begin{table}[!ht]
\centering
\caption{Integration result for $\phi_{0}\left(k=2\right)=\frac{\pi}{3}=60\text{\textdegree}$ integrated until $t=1$ (using an order 20 Picard-iteration based integrator with stepsize $h=10^{-3}$ over 1000 iterations within COSY Infinity). The component $\mathcal{M}_{0}^{+}=Q\left(q_{0},p_{0}\right)$ is on the left, $\mathcal{M}_{0}^{-}=P\left(q_{0},p_{0}\right)$ on the right.}
\label{tab:M0}
\vspace{10pt}
\begin{tabular}{|c||c|r||c|r|}
\hline 
O & Coeff. & Value & Coeff. & Value\\
\hline 
\hline 
0 & $q_{const}$ & 1.04719755 & $p_{const}$ & 0\\
\hline \hline
1 & $\left(Q|q_{0}\right)$ & 0.33918599 & $\left(P|q_{0}\right)$ & -1.15214118\\
\hline 
1 & $\left(Q|p_{0}\right)$ & 0.76809412 & $\left(P|p_{0}\right)$ & 0.33918599\\
\hline \hline
2 & $\mathcal{U}_{2(2,0)}^{+}$ & -0.44622446 & $\mathcal{U}_{2(2,0)}^{-}$ & -0.55821731\\
\hline 
2 & $\mathcal{U}_{2(1,1)}^{+}$ & -0.29304415 & $\mathcal{U}_{2(1,1)}^{-}$ & -0.64033440\\
\hline 
2 & $\mathcal{U}_{2(0,2)}^{+}$ & -0.08403817 & $\mathcal{U}_{2(0,2)}^{-}$ & -0.29304415\\
\hline \hline
3 & $\mathcal{U}_{3(3,0)}^{+}$ & 0.31844278 & $\mathcal{U}_{3(3,0)}^{-}$ & 0.50817317\\
\hline 
3 & $\mathcal{U}_{3(2,1)}^{+}$ & 0.29904862 & $\mathcal{U}_{3(2,1)}^{-}$ & 0.76091921\\
\hline 
3 & $\mathcal{U}_{3(1,2)}^{+}$ & 0.13758223 & $\mathcal{U}_{3(1,2)}^{-}$ & 0.46230241\\
\hline 
3 & $\mathcal{U}_{3(0,3)}^{+}$ & 0.03017663 & $\mathcal{U}_{3(0,3)}^{-}$ & 0.13758223\\
\hline 
\end{tabular}
\end{table}

For this specific example, the calculation of the fixed point map is straight forward from the flow and works for any time $t$, since the equilibrium /fixed point does not change over time for a fixed $\omega$. However, for higher dimensional systems e.g. accelerators, it is often necessary to consider the motion within a Poincaré surface (e.g. a vertical cross-section) with a fixed point Poincaré return map in order to obtain a fixed point map. In Ref. \cite{grote2006} the calculation of a Poincaré map using a timewise projection is explained.

\section{DA normal form algorithm}

Before diving into the steps of the DA normal form algorithm we are clarifying the notation. The map is dependent on the \lq original\rq\, variables $(q_{0},p_{0})$. The transformations of the algorithm are done order-by-order. With each transformation, the index of the map and the variables is going to increase by 1, i.e. as a result of the first (order) transformation (the linear transformation) we get $\mathcal{M}_{1}$ dependent on the variables $(q_{1},p_{1})$. For each order $m$ there is a transformation $\mathcal{A}_{m}$ and its inverse $\mathcal{A}_{m}^{-1}$, which are applied to resulting map of the previous transformation $\mathcal{M}_{m-1}$ to yield the resulting map of the $m$th order transformation $\mathcal{M}_{m}(q_{m},p_{m})=(\mathcal{A}_{m}\circ\mathcal{M}_{m-1}\circ\mathcal{A}_{m}^{-1})(q_{m},p_{m})$. Note that $\mathcal{A}_{m}^{-1}$ transforms $(q_{m},p_{m})$ to $(q_{m-1},p_{m-1})$, which are the variables of the map of the previous order $\mathcal{M}_{m-1}$. $\mathcal{A}_{m}$ transforms the intermediate result of $\mathcal{M}_{m-1}\circ\mathcal{A}_{m}^{-1}$, which is in the $(q_{m-1},p_{m-1})$-phase space back to the new phase space in $(q_{m},p_{m})$. To express maps which are not always associated with the same variables we use the variables $(q,p)$ without an index. In Ref. \cite{Berz1999MMM}, the variables $(q_{m},p_{m})$ are denoted with the $(s^{+},s^{-})$ notation and the normal form coordinates $(q_\text{NF},p_\text{NF})$ are written as $(t^{+},t^{-})$ instead.

\subsection{(Parameter dependent) fixed point}

The DA normal form algorithm starts with an origin preserving map. Accordingly, the result from the integration is shifted to the equilibrium/fixed point $\mathcal{M}_{FP}=\mathcal{M}_{0}-\mathcal{C}$, hence $\mathcal{M}_{FP}=\mathcal{L}+\sum_{m}\mathcal{U}_{m}$ is an origin preserving fixed point map with $\mathcal{M}_{FP}(\vec{0})=\vec{0}$. 

If the map were parameter dependent, e.g. on changes in the driving frequency $\omega=\omega_{0}+\delta\omega$, the normal form algorithm would require the calculation of the parameter dependent fixed point $\vec{z}(\delta\omega)=(q_{FP}(\delta\omega),p_{FP}(\delta\omega))$ such that $\mathcal{M}_{FP}(\vec{0},\delta\omega)=\vec{0}$. In eq. \ref{eq:equipoint}, the relation of the equilibrium / fixed point and the driving frequency was already calculated yielding the parameter dependent fixed point
\begin{equation}
\vec{z}\left(\delta\omega\right)=\left(\arccos\left(\frac{g}{\left(\omega_{0}+\delta\omega\right)^{2}R}\right),0\right)\quad\text{for}\quad\left| \frac{g}{\left(\omega_{0}+\delta\omega\right)^{2}R} \right| \le 1.\nonumber
\end{equation}

For less straight forward systems, one uses the following inversion method on the extended map $(\mathcal{M}_{FP}-\mathcal{I}_{\vec{z}},\mathcal{I}_{\vec{\eta}})$ to find the parameter dependent fixed point $\vec{z}(\vec{\eta})$ \cite[eq. 7.47]{Berz1999MMM}:
\begin{equation}
\left(\vec{z}\left(\vec{\eta}\right),\mathcal{I}_{\vec{\eta}}\right)=\left(\mathcal{M}_{FP}-\mathcal{I}_{\vec{z}},\mathcal{I}_{\vec{\eta}}\right)^{-1}\left(\vec{0},\vec{\eta}\right),\label{eq:PDFP1}
\end{equation}
where $\mathcal{I}_{\vec{z}}$ and $\mathcal{I}_{\vec{\eta}}$ are the identity map of $\vec{z}$ and $\vec{\eta}$, respectively. 

Given the parameter dependent fixed point, the map is expanded around it:
\begin{equation}
\mathcal{M}_{PDFP}=\mathcal{M}_{FP}\left(\vec{z}\left(\vec{\eta}\right)+\vec{z},\vec{\eta}\right)-\mathcal{M}_{FP}\left(\vec{z}\left(\vec{\eta}\right),\vec{\eta}\right).\label{eq:PDFP2}
\end{equation}

This introduction does not consider parameter dependence, which is why the example calculation is proceeding with $\mathcal{M}_{FP}$.

\subsection{Linear transformation}

The first order transformation is the diagonalization, transforming the system into the eigenvector space of the linear part $\mathcal{L}$. In order to determine the transformation $\mathcal{A}_{1}$ and its inverse $\mathcal{A}_{1}^{-1}$ for the diagonalization, we determine the eigenvalues $\lambda_{\pm}$ and eigenvectors $\vec{v}_{\pm}$ of the linear matrix $\hat{L}$ in the linear part $\mathcal{L}$. For this we require that all eigenvalues of $\mathcal{M}_{FP}$ are distinct. Furthermore, we only consider cases where $\mathcal{M}_{FP}$ is linearly stable, which means that all eigenvalues have an absolute value $|\lambda|\le1$. This also means that $\det(\hat{L})\le1$, otherwise at least one of the eigenvalues is larger than 1, making the system linearly unstable. Particularly interesting is the case $\det(\hat{L})=1$, which indicates that the system is symplectic and only stable in the case of complex conjugate eigenvalues$\lambda_{\pm}=e^{\pm i\mu}$. While there are procedures for the cases of real and degenerate eigenvalues with magnitude smaller than one (see Ref. \cite{Berz1999MMM}), this work only illustrates the procedures for the most relevant and common symplectic case of only complex conjugate eigenvalues and eigenvectors. 

Solving the characteristic polynomial yields the eigenvalues
\begin{align}
\lambda_{\pm} & =\frac{\text{tr}\left(\hat{L}\right)}{2}\pm\sqrt{\frac{\text{tr}\left(\hat{L}\right)^{2}}{4}-\det\left(\hat{L}\right)}=re^{\pm i\mu}\nonumber\\
\text{with}\quad r & =\sqrt{\det\left(\hat{L}\right)}\quad\text{and}\quad\mu=\text{sign}\left(Q|p_{0}\right)\arccos\left(\frac{\text{tr}\left(\hat{L}\right)}{2r}\right).\nonumber
\end{align}

To generalize the procedure of diagonalization, the Twiss parameters\cite{Courant1958TwissPara} are used with
\begin{equation}
\alpha=\frac{(Q|q_{0})-(P|p_{0})}{2r\sin\mu}\quad\beta=\frac{(Q|p_{0})}{r\sin\mu}\quad\gamma=\frac{-(P|q_{0})}{r\sin\mu}.\nonumber
\end{equation}

With this notation the linear matrix $\hat{L}$ can be generally written as 
\begin{equation}
\hat{L}=\begin{pmatrix}\cos\mu+\alpha\sin\mu & \beta\sin\mu\\
-\gamma\sin\mu & \cos\mu-\alpha\sin\mu
\end{pmatrix}.\nonumber
\end{equation}

The complex conjugate eigenvectors $\vec{v}_{\pm}$ associated with the complex conjugate eigenvalues $\lambda_{\pm}$ of $\hat{L}$ are then obtained by solving  
$\left(\hat{L}-\lambda_{\pm}\mathcal{I}\right)\vec{v}_{\pm}=\vec{0}.$ 

As a result, the following eigenvectors are calculated
\begin{equation}
\vec{v}_{\pm}=\rho e^{i\eta}\begin{pmatrix}-\beta\\
\alpha\mp i
\end{pmatrix}\quad\text{or}\quad\vec{v}_{\pm}=\rho e^{i\eta}\begin{pmatrix}\alpha\pm i\\
\gamma
\end{pmatrix},\nonumber
\end{equation}
for the case that either $\beta=0$ or $\gamma=0$. The eigenvectors have a magnitude-freedom $\rho$ and phase-freedom $\eta$. The magnitude is chosen such that the transformation is non-scaling $\left|\det\left(\mathcal{A}_{1}\right)\right|=1$, which is particularly important for the transformation of the nonlinear terms. Since the new coordinates are complex conjugate pairs $\left(q_{1},p_{1}\right)$ such that $\bar{q}_{1}=p_{1}$, the phase of the eigenvectors are chosen such that the transformation $\mathcal{A}_{1}^{-1}$ consist of the two complex conjugate eigenvectors $\vec{v}_{\pm}$, guaranteeing that $\mathcal{A}_{1}^{-1}\left(q_{1},p_{1}\right)$ is real just like the original variables $\left(q_0,p_0\right)$ and the fixed point map $\mathcal{M}_{FP}$. Furthermore, $\mathcal{A}_{1}$ is chosen such that the resulting map $\mathcal{M}_{1}=\mathcal{A}_{1}\circ\mathcal{M}_{FP}\circ\mathcal{A}_{1}^{-1}$ is in the complex conjugate eigenvector space and has complex conjugate components $\bar{\mathcal{M}}_{1}^{+}=\mathcal{M}_{1}^{-}$. Accordingly, the transformation $\mathcal{A}_{1}$ and its inverse $\mathcal{A}_{1}^{-1}$ (for $\beta\ne0$) are given by 
\begin{eqnarray}
\mathcal{A}_{1}^{-1} & = & \begin{pmatrix}(q_{0}|q_{1}) & (q_{0}|p_{1})\\
(p_{0}|q_{1}) & (p_{0}|p_{1})
\end{pmatrix}=\frac{1}{\sqrt{2\beta}}\begin{pmatrix}-\beta & -\beta\\
\alpha-i & \alpha+i
\end{pmatrix}\label{eq:A1-1}\\
\mathcal{A}_{1} & = & \begin{pmatrix}(q_{1}|q_{0}) & (q_{1}|p_{0})\\
(p_{1}|q_{0}) & (p_{1}|p_{0})
\end{pmatrix}=\frac{i}{\sqrt{2\beta}}\begin{pmatrix}\alpha+i & +\beta\\
i-\alpha & -\beta
\end{pmatrix}.\label{eq:A1}
\end{eqnarray}

For the centrifugal governor example with $k=2$, the eigenvalues are $\lambda_{\pm}=re^{\pm i\mu}$ with $r=1$ and $\mu=1.22474487$. The Twiss parameters are
\begin{equation}
\alpha=0\quad\beta=0.816496581\quad\gamma=1.22474487,\nonumber
\end{equation}
which yields the following transformation matrices from eq. \ref{eq:A1-1} and \ref{eq:A1}:
\begin{eqnarray}
\mathcal{A}_{1}^{-1}=\begin{pmatrix}-0.638943104 & -0.638943104\\
-0.782542290\,i & 0.782542290\,i
\end{pmatrix}\nonumber\\
\mathcal{A}_{1}	=\begin{pmatrix}-0.782542290 & 0.638943104\,i\\
-0.782542290 & -0.638943104\,i
\end{pmatrix}.\nonumber
\end{eqnarray}

The resulting diagonalized map is of the form $\mathcal{M}_{1}=\mathcal{R}+\sum_{m}S_{m}$, where $\mathcal{S}_{m}$ are the transformed nonlinear parts of order $m$ in the eigenvector space of $\hat{L}$ and $\mathcal{R}$ is the diagonalized linear part, where the linear matrix $\hat{R}$ of $\mathcal{R}$ only consist of the eigenvalues $e^{\pm i\mu}$ on its main diagonal:
\begin{eqnarray}
\mathcal{M}_{1}\left(q_{1},p_{1}\right) & = & \underbrace{\begin{pmatrix}e^{i\mu} & 0\\
0 & e^{-i\mu}
\end{pmatrix}\begin{pmatrix}q_{1}\\
p_{1}
\end{pmatrix}}_{\mathcal{R}}\nonumber\\
 & + & \underbrace{\begin{pmatrix}\mathcal{S}_{2(2,0)}^{+}\\
\mathcal{S}_{2(2,0)}^{-}
\end{pmatrix}q_{1}^{2}+\begin{pmatrix}\mathcal{S}_{2(1,1)}^{+}\\
\mathcal{S}_{2(1,1)}^{-}
\end{pmatrix}q_{1}p_{1}+\begin{pmatrix}\mathcal{S}_{2(0,2)}^{+}\\
\mathcal{S}_{2(0,2)}^{-}
\end{pmatrix}p_{1}^{2}}_{\mathcal{S}_{2}}\nonumber\\
 & + & \underbrace{\begin{pmatrix}\mathcal{S}_{3(3,0)}^{+}\\
\mathcal{S}_{3(3,0)}^{-}
\end{pmatrix}q_{1}^{3}+\begin{pmatrix}\mathcal{S}_{3(2,1)}^{+}\\
\mathcal{S}_{3(2,1)}^{-}
\end{pmatrix}q_{1}^{2}p_{1}+...}_{\mathcal{S}_{3}}\label{eq:M1}
\end{eqnarray}
Table \ref{tab:M1} lists the values to the coefficients above for the centrifugal governor example case.
\begin{table}[!ht]
\caption{Coefficients of $\mathcal{M}_{1}$ up to order 3. Note the complex conjugate property $\mathcal{S}_{m\left(k_{+},k_{-}\right)}^{\pm}=\mathcal{\bar{S}}_{m\left(k_{-},k_{+}\right)}^{\mp}$. The component $\mathcal{M}_{1}^{+}$ is on the left, $\mathcal{M}_{1}^{-}$ on the right.}
\begin{centering}
\begin{tabular}{|c||c|r|r||c|r|r|}
\hline 
O & Coeff. & Real Part & Imaginary Part & Coeff. & Real Part & Imaginary Part\\
\hline 
\hline 
1 & $\text{e}^{i\mu}$ & 0.339185989 & 0.940719334 & $\text{e}^{-i\mu}$ & 0.339185989 & -0.940719334\\
\hline \hline
2 & $\mathcal{S}_{2(2,0)}^{+}$ & 0.306852938  & 0.083709890  & $\mathcal{S}_{2(2,0)}^{-}$ & -0.102284313  & 0.145609550 \\
\hline 
2 & $\mathcal{S}_{2(1,1)}^{+}$ & 0.365655459  & -0.520538539  & $\mathcal{S}_{2(1,1)}^{-}$ & 0.365655459  & 0.520538539 \\
\hline 
2 & $\mathcal{S}_{2(0,2)}^{+}$ & -0.102284313  & -0.145609550  & $\mathcal{S}_{2(0,2)}^{-}$ & 0.306852938  & -0.083709890 \\
\hline \hline
3 & $\mathcal{S}_{3(3,0)}^{+}$ & 0.136072276  & 0.094325994  & $\mathcal{S}_{3(3,0)}^{-}$ & -0.090320124  & 0.032565845 \\
\hline 
3 & $\mathcal{S}_{3(2,1)}^{+}$ & 0.518830697  & -0.260951321  & $\mathcal{S}_{3(2,1)}^{-}$ & -0.044567972  & 0.478373053 \\
\hline 
3 & $\mathcal{S}_{3(1,2)}^{+}$ & -0.044567972  & -0.478373053  & $\mathcal{S}_{3(1,2)}^{-}$ & 0.518830697  & 0.260951321 \\
\hline 
3 & $\mathcal{S}_{3(0,3)}^{+}$ & -0.090320124  & -0.032565845  & $\mathcal{S}_{3(0,3)}^{-}$ & 0.136072276  & -0.094325994 \\
\hline 
\end{tabular}
\par\end{centering}
\centering{}\label{tab:M1}
\end{table}

\subsection{Nonlinear transformations\label{subsec:Non-linear-Transformations}}

The nonlinear transformations are the key part of the normal form algorithm. In this first part of this subsection we are going look at an $m$th order transformation in general, before going through the nonlinear transformation for order two and three in detail. Don't get discouraged from reading further if the general notation is a bit overwhelming. 

\subsubsection{General mth order nonlinear transformation}

All the following nonlinear transformations are done order-by-order and are all of the same form: $\mathcal{M}_{m}=\mathcal{A}_{m}\circ\mathcal{M}_{m-1}\circ\mathcal{A}_{m}^{-1}$, where the $m$th transformation does \textbf{not} change any of the lower order terms of $\mathcal{M}_{m-1}$ that have already been transformed in the previous transformations. Hence, $\mathcal{M}_{m}$ differs from $\mathcal{M}_{m-1}$ only in the orders $m$ and larger. The $m$th order transformation $\mathcal{A}_{m}=\mathcal{I}+\mathcal{T}_{m}+\mathcal{O}_{\ge m+1}$, specifically the polynomial $\mathcal{T}_{m}$ of only $m$th order terms, is chosen such that the $m$th order terms $\mathcal{S}_m$ of the map $\mathcal{M}_{m-1}$ are simplified or even eliminated. Effects on the higher orders due to the $m$th order transformation can only be considered by adjusting the terms of order higher than $m$, namely $\mathcal{O}_{\ge m+1}$. In other words, finding $\mathcal{T}_{m}$ is essential to the DA normal form algorithm, the terms $\mathcal{O}_{\ge m+1}$ on the other hand can be chosen freely i.e. to make the transformation symplectic $\left(\mathcal{A}_{m}=\exp\left(L_{\mathcal{T}_{m}}\right)\right)$ or to avoid higher order resonances. Usually, the symplectic transformation is chosen since the calculation of the transformation $\mathcal{A}_{m}$ and its inverse are straight forward. 

The flow operator $L_{\mathcal{T}_{m}}=(\mathcal{T}_m^+\partial_q+\mathcal{T}_m^-\partial_p)$ in the exponential behaves in the following way:
\begin{eqnarray}
	\exp\left(L_{\mathcal{T}_{m}}\right)\mathcal{I}&=&\left( L_{\mathcal{T}_{m}}^0+L_{\mathcal{T}_{m}}^1+\frac{1}{2}L_{\mathcal{T}_{m}}^2+\mathcal{O}_{>(m+1)}\right)\mathcal{I}\nonumber\\
	&=& \left(1+(\mathcal{T}_m^+\partial_q+\mathcal{T}_m^-\partial_p)+\frac{1}{2}L_{\mathcal{T}_{m}}(\mathcal{T}_m^+\partial_q+\mathcal{T}_m^-\partial_p)+\mathcal{O}_{>(m+1)}\right)\begin{pmatrix}q\\p\end{pmatrix}\nonumber\\
	&=& \mathcal{I} + \mathcal{T}_m + \frac{1}{2}L_{\mathcal{T}_{m}}\mathcal{T}_m +\mathcal{O}_{>(m+1)}.
\end{eqnarray}
Accordingly, the inverse is given by
\begin{equation}
\mathcal{A}_{m}^{-1}=\exp\left(-L_{\mathcal{T}_{m}}\right)=\mathcal{I} - \mathcal{T}_m + \frac{1}{2}L_{\mathcal{T}_{m}}\mathcal{T}_m - \mathcal{O}_{>(m+1)}.
\end{equation}

In the example case of the centrifugal governor we investigate the DA normal form algorithm up to order three, which means for $m=3$:
\begin{align}
\mathcal{A}_{3} & =\exp\left(\quad L_{\mathcal{T}_{3}}\right)\mathcal{I}=_3 \mathcal{I}+\mathcal{T}_{3}\\
\mathcal{A}_{3}^{-1} & =\exp\left(-L_{\mathcal{T}_{3}}\right)\mathcal{I}=_{3}\;\mathcal{I}-\mathcal{T}_{3}.
\end{align}
For the second order transformation it is necessary to consider the third order terms $\mathcal{O}_3$, since they influence the third order terms of $\mathcal{M}_2$:
\begin{align}
\mathcal{A}_{2} & =\exp\left(\quad L_{\mathcal{T}_{2}}\right)\mathcal{I}=_3 \mathcal{I}+\mathcal{T}_{2}+\mathcal{O}_3\label{eq:A2}\\
\mathcal{A}_{2}^{-1} & =\exp\left(-L_{\mathcal{T}_{2}}\right)\mathcal{I}=_3\;\mathcal{I}-\mathcal{T}_{2}+\mathcal{O}_3\label{eq:A2-1}
\end{align}
with
\begin{eqnarray}
\mathcal{O}_3=\frac{1}{2}L_{\mathcal{T}_{m}}\mathcal{T}_m= \frac{1}{2}(\mathcal{T}_2^+\partial_q+\mathcal{T}_2^-\partial_p)\mathcal{T}_2.\label{eq:O3}
\end{eqnarray}
The '$=_{m}$'-notation indicates that the quantities on both sides are equal up to expansion order $m$. 
 
In order to determine $\mathcal{T}_{m}$, we analyze the $m$th order transformation and only look at terms up to order $m$ \cite[eq. 7.62]{Berz1999MMM}: 
\begin{eqnarray}
\mathcal{A}_{m}\circ\mathcal{M}_{m-1}\circ\mathcal{A}_{m}^{-1} & =_{m} & \left(\mathcal{I}+\mathcal{T}_{m}\right)\circ\left(\mathcal{R}+\mathcal{S}_{m}\right)\circ\left(\mathcal{I}-\mathcal{T}_{m}\right)\nonumber\\
 & =_{m} & \left(\mathcal{I}+\mathcal{T}_{m}\right)\circ\left(\mathcal{R}-\mathcal{R}\circ\mathcal{T}_{m}+\mathcal{S}_{m}\right)\nonumber\\
 & =_{m} & \mathcal{R}+\mathcal{S}_{m}+\left[\mathcal{T}_{m},\mathcal{R}\right].\label{eq:aMa3}
\end{eqnarray}

Note that various terms with orders higher than $m$ are ignored in the equations above. The goal is to choose $\mathcal{T}_{m}$ such that the commutator $\left[\mathcal{T}_{m},\mathcal{R}\right]=\mathcal{T}_{m}\circ\mathcal{R}-\mathcal{R}\circ\mathcal{T}_{m}=-\mathcal{S}_{m}$ to simplify $\mathcal{M}_m$, i.e. the result of eq. \ref{eq:aMa3}. The polynomials in the upper and lower component of $\mathcal{T}_{m}$ can be express as
\begin{equation}
\mathcal{T}_{m}^{\pm}\left(q,p\right)=\sum_{\begin{array}{c}
k_{+}+k_{-}=m\\
k_{\pm}\in\mathbb{\mathbb{N}}_{0}
\end{array}}\mathcal{T}_{m(k_{+},k_{-})}^{\pm}q^{k_{+}}p^{k_{-}}.\label{eq:Tm}
\end{equation}
Accordingly, the commutator $\mathcal{C}_{m}=\left[\mathcal{T}_{m},\mathcal{R}\right]$ yields
\begin{equation}
\mathcal{C}_{m}^{\pm}\left(q,p\right)=\sum_{\begin{array}{c}
k_{+}+k_{-}=m\\
k_{\pm}\in\mathbb{\mathbb{N}}_{0}
\end{array}}\mathcal{T}_{m(k_{+},k_{-})}^{\pm}\left(e^{i\mu\left(k_{+}-k_{-}\right)}-e^{\pm i\mu}\right)q^{k_{+}}p^{k_{-}}.
\end{equation}

A term in $\mathcal{S}_m$ can only be removed if and only if the corresponding term in the commutator $\mathcal{C}_{m}$ is not zero. Terms of the commutator are zero, whenever the condition 
\begin{equation}
e^{i\mu\left(k_{+}-k_{-}\right)}-e^{\pm i\mu}=0\label{eq:NFcondition}
\end{equation}
is satisfied, which is the case for $k_{+}-k_{-}=\pm1$. This is the key condition (eq. \ref{eq:NFcondition}) of the DA normal form algorithm, since it determines the surviving nonlinear terms $\mathcal{S}_m$. All other terms that do not satisfy the condition are eliminated by choosing the coefficients of $\mathcal{T}_{m}$ as follows
\begin{equation}
\mathcal{T}_{m(k_{+},k_{-})}^{\pm}=\frac{-S_{m(k_{+},k_{-})}^{\pm}}{e^{i\mu\left(k_{+}-k_{-}\right)}-e^{\pm i\mu}}.
\end{equation}
Specifically, this means that the terms $\mathcal{S}_{m(k,k-1)}^{+}$ and $\mathcal{S}_{m(k-1,k)}^{-}$ always survive for all uneven orders $m$ with $m=k+k-1=2k-1$. 

\subsubsection{Explicit second order nonlinear transformation}

The polynomial $\mathcal{T}_{m}$ from eq. \ref{eq:Tm} for $m=2$ yields
\begin{align}
\mathcal{T}_{2}\left(q,p\right) & =\left(\mathcal{T}_{2}^{\pm}|2,0\right)q^{2}+\left(\mathcal{T}_{2}^{\pm}|1,1\right)qp+\left(\mathcal{T}_{2}^{\pm}|0,2\right)p^{2}\nonumber\\
 & =\begin{pmatrix}\mathcal{T}_{2(2,0)}^{+}\\
\mathcal{T}_{2(2,0)}^{-}
\end{pmatrix}q^{2}+\begin{pmatrix}\mathcal{T}_{2(1,1)}^{+}\\
\mathcal{T}_{2(1,1)}^{-}
\end{pmatrix}qp+\begin{pmatrix}\mathcal{T}_{2(0,2)}^{+}\\
\mathcal{T}_{2(\text{0},2)}^{-}
\end{pmatrix}p^{2}.
\end{align}

The commutator $\mathcal{C}_{2}=\left[\mathcal{T}_{2},\mathcal{R}\right]=\mathcal{T}_{2}\circ\mathcal{R}-\mathcal{R}\circ\mathcal{T}_{2}$ of the second (even) order nonlinear transformation has only non-zero terms with
\begin{align}
\mathcal{C}_{2}\left(q,p\right) & =\left[\mathcal{T}_{2},\mathcal{R}\right]\left(q,p\right)=\left(\mathcal{T}_{2}\circ\mathcal{R}-\mathcal{R}\circ\mathcal{T}_{2}\right)\left(q,p\right)\nonumber\\
 & =\left(\mathcal{T}_{2}^{\pm}|2,0\right)e^{2i\mu}q^{2}+\left(\mathcal{T}_{2}^{\pm}|1,1\right)qp+\left(\mathcal{T}_{2}^{\pm}|0,2\right)e^{-2i\mu}p^{2}-e^{\pm i\mu}\mathcal{T}_{2}^{\pm}\left(q,p\right)\nonumber\\
 & =\begin{pmatrix}\mathcal{T}_{2(2,0)}^{+}\left(e^{2i\mu}-e^{i\mu}\right)\\
\mathcal{T}_{2(2,0)}^{-}\left(e^{2i\mu}-e^{-i\mu}\right)
\end{pmatrix}q^{2}+\begin{pmatrix}-e^{i\mu}\mathcal{T}_{2(1,1)}^{+}\\
-e^{-i\mu}\mathcal{T}_{2(1,1)}^{-}
\end{pmatrix}qp+\begin{pmatrix}\mathcal{T}_{2(0,2)}^{+}\left(e^{-2i\mu}-e^{i\mu}\right)\\
\mathcal{T}_{2(\text{0},2)}^{-}\left(e^{-2i\mu}-e^{-i\mu}\right)
\end{pmatrix}p^{2},
\end{align}
eliminating all $\mathcal{S}_{2}$ terms by choosing
\begin{equation}
\mathcal{T}_{2(k_{+},k_{-})}^{\pm}=\frac{-S_{2(k_{+},k_{-})}^{\pm}}{\left(e^{i\mu\left(k_{+}-k_{-}\right)}-e^{\pm i\mu}\right)},\label{eq:T2(S2)}
\end{equation}
since the condition from eq. \ref{eq:NFcondition} is not satisfied:
\begin{equation}
e^{i\mu\left(k_{+}-k_{-}\right)}-e^{\pm i\mu}\ne0\quad\forall k_{+},k_{-}\in\mathbb{N}_{0}\quad\text{with}\quad k_{+}+k_{-}=2.\nonumber
\end{equation}

The values of the $\mathcal{T}_{2(k_{+},k_{-})}^{\pm}$ for the centrifugal governor example are given in table \ref{tab:T2O3}. The terms of $\mathcal{O}_3$ are calculated via eq. \ref{eq:O3} from $\mathcal{T}_2$ and are also given in table \ref{tab:T2O3} yielding all terms of the transformation $\mathcal{A}_2$ and its inverse $\mathcal{A}_2^{-1}$ from eq. \ref{eq:A2} and \ref{eq:A2-1}.

\begin{table}[!ht]
\centering
\caption{The values of the $\mathcal{T}_{2(k_{+},k_{-})}^{\pm}$ and $\mathcal{O}_{3(k_{+},k_{-})}^{\pm}$. Note that $\mathcal{T}_2$ and $\mathcal{O}_3$ and therefore $\mathcal{A}_2$ and its inverse are real with $\mathcal{A}_{m(k_{+},k_{-})}^{+}=\mathcal{A}_{m(k_{-},k_{+})}^{-}$.}
\label{tab:T2O3}
\begin{tabular}{|c||c|r||c|r|}
\hline 
O &Coeff. & Value & Coeff. & Value\\
\hline 
\hline 
2 & $\mathcal{T}_{2(2,0)}^{+}$ & 0.276670480 & $\mathcal{T}_{2(2,0)}^{-}$ & -0.092223493\\
\hline 
2 & $\mathcal{T}_{2(1,1)}^{+}$ &-0.553340960 & $\mathcal{T}_{2(1,1)}^{-}$ & -0.553340960\\
\hline 
2 & $\mathcal{T}_{2(0,2)}^{+}$ &-0.092223493 & $\mathcal{T}_{2(0,2)}^{-}$ &  0.276670480
\\
\hline 
\hline 
3 & $\mathcal{O}_{3(3,0)}^{+}$ & 0.102062073 & $\mathcal{O}_{3(3,0)}^{-}$ & 0\\
\hline 
3 & $\mathcal{O}_{3(2,1)}^{+}$ &-0.068041382 & $\mathcal{O}_{3(2,1)}^{-}$ & 0.102062073\\
\hline 
3 & $\mathcal{O}_{3(1,2)}^{+}$ &0.102062073 & $\mathcal{O}_{3(1,2)}^{-}$ &  -0.068041382\\
\hline
3 & $\mathcal{O}_{3(0,3)}^{+}$ & 0 & $\mathcal{O}_{3(0,3)}^{-}$ &  0.102062073\\
\hline 
\end{tabular}
\end{table}

To study how the second order transformation affects the third order terms $\mathcal{S}_3$ of the map 
$\mathcal{M}_2$, the transformation is considered up to third order: 
\begin{eqnarray}
\mathcal{M}_{2} & =_{3} & \mathcal{A}_{2}\circ\mathcal{M}_{1}\circ\mathcal{A}_{2}^{-1}\nonumber\\
 & =_{3} & \left(\mathcal{I}+\mathcal{T}_{2}+\mathcal{O}_{3}\right)\circ\left(\mathcal{R}+\mathcal{S}_{2}+\mathcal{S}_{3}\right)\circ\left(\mathcal{I}-\mathcal{T}_{2}+\mathcal{O}_{3}\right)\nonumber\\
 & =_{3} & \left(\mathcal{I}+\mathcal{T}_{2}+\mathcal{O}_{3}\right)\circ\left(\mathcal{R}\circ\left(\mathcal{I}-\mathcal{T}_{2}+\mathcal{O}_{3}\right)+\underline{\mathcal{S}_{2}\circ\left(\mathcal{I}-\mathcal{T}_{2}+\mathcal{O}_{3}\right)}+\underline{\mathcal{S}_{3}\circ\left(\mathcal{I}-\mathcal{T}_{2}+\mathcal{O}_{3}\right)}\right)\nonumber\\
 & =_{3} & \left(\mathcal{I}+\mathcal{T}_{2}+\mathcal{O}_{3}\right)\circ\left(\mathcal{R}-\mathcal{R}\circ\mathcal{T}_{2}+\mathcal{R}\circ\mathcal{O}_{3}+\overbrace{\mathcal{S}_{2}+\mathcal{S}_{2\rightarrow3}+\cancel{\mathcal{O}_{\ge 4}}}+\overbrace{\mathcal{S}_{3}+\cancel{\mathcal{O}_{\ge 4}}}\right)\nonumber\\
 & =_{3} & \mathcal{R}-\mathcal{R}\circ\mathcal{T}_{2}+\mathcal{R}\circ\mathcal{O}_{3}+\mathcal{S}_{2}+\mathcal{S}_{2\rightarrow3}+\mathcal{S}_{3}\nonumber\\
 & + & \mathcal{T}_{2}\circ\underline{\left(\mathcal{R}-\mathcal{R}\circ\mathcal{T}_{2}+\mathcal{R}\circ\mathcal{O}_{3}+\mathcal{S}_{2}+\mathcal{S}{}_{2\rightarrow3}+\mathcal{S}_{3}\right)}+\cancel{\mathcal{O}_{\ge 4}}\nonumber\\
 & =_{3} & \overbrace{\mathcal{T}_{2}\circ\mathcal{R}+\mathcal{K}_{2\rightarrow3}+\cancel{\mathcal{O}_{\ge 4}}}+\mathcal{R}-\mathcal{R}\circ\mathcal{T}_{2}+\mathcal{R}\circ\mathcal{O}_{3}+\mathcal{S}_{2}+\mathcal{S}_{2\rightarrow3}+\mathcal{S}_{3}\nonumber\\
 & =_{3} & \mathcal{R}+\underbrace{\mathcal{S}_{2}+[\mathcal{T}_{2}\circ\mathcal{R}]}_{=0}+\underbrace{\mathcal{S}_{3}+\mathcal{S}_{2\rightarrow3}+\mathcal{K}_{2\rightarrow3}+\mathcal{R}\circ\mathcal{O}_{3}}_{\mathcal{S}_{3,new}}
\end{eqnarray}

All the crossed-out terms $\cancel{\mathcal{O}_{\ge 4}}$ represent terms that do not contribute to the result up to order three, since they are at least of order four. As a result of the second order transformation there are three new terms of order 3: $\mathcal{S}_{2\rightarrow3}=_{3}\mathcal{S}_{2}\circ\left(\mathcal{I}-\mathcal{T}_{2}\right)-\mathcal{S}_{2}$, $\mathcal{K}_{2\rightarrow3}=_{3}\mathcal{T}_{2}\circ\left(\mathcal{R}-\mathcal{R}\circ\mathcal{T}_{2}+\mathcal{S}_{2}\right)-\mathcal{T}_{2}\circ\mathcal{R}$ and $\mathcal{R}\circ\mathcal{O}_{3}$. The first two are calculated more explicitly in the appendix \ref{sec:Appendix}. The result of the second order transformation $\mathcal{M}_{2}=\mathcal{R}+\mathcal{S}_{3,new}$ for the example case of the centrifugal governor is given in table \ref{tab:M2}.

\begin{table}[!ht]
\centering
\caption{New coefficients of third order of $\mathcal{M}_{2}$ after the second order transformation. Note that the first order terms remain unchanged and that the second order terms are all eliminated by the second order transformation. Interestingly, the second order transformation caused terms of the third order to disappear in this specific case, this is \textbf{not} a general property of the second order transformation. The component $\mathcal{M}_{2}^{+}$ is on the left, $\mathcal{M}_{2}^{-}$ on the right. The emphasized terms are surviving the third order transformation as explained in the following subsection.}
\label{tab:M2}
\begin{tabular}{|c||c|r|r||c|r|r|}
\hline 
O & Coeff. & Real Part & Imaginary Part & Coeff. & Real Part & Imaginary Part\\
\hline \hline
3 & $\mathcal{S}_{3,new(3,0)}^{+}$ & 0.122541282  & 0.147840016  & $\mathcal{S}_{3,new(3,0)}^{-}$ & 0 & 0\\
\hline 
3 & $\mathcal{S}_{3,new(2,1)}^{+}$ & \textbf{0.940719334} & \textbf{-0.339185989} & $\mathcal{S}_{3,new(2,1)}^{-}$ & 0 & 0.576070590 \\
\hline 
3 & $\mathcal{S}_{3,new(1,2)}^{+}$ & 0 & -0.576070590  & $\mathcal{S}_{3,new(1,2)}^{-}$ & \textbf{0.940719334} & \textbf{0.339185989}\\
\hline 
3 & $\mathcal{S}_{3,new(0,3)}^{+}$ & 0 & 0 & $\mathcal{S}_{3,new(0,3)}^{-}$ & 0.122541282 & -0.147840016\\
\hline 
\end{tabular}
\end{table}

\subsubsection{Explicit third order nonlinear transformation}

The third order transformation follows the same scheme as above (eq. \ref{eq:aMa3}) only that the commutator $\mathcal{C}_{3}=\left[\mathcal{T}_{3}\circ\mathcal{R}\right]$ has terms that are zero
\begin{align}
\mathcal{C}_{3} & =\begin{pmatrix}\mathcal{T}_{3(3,0)}^{+}\left(e^{3i\mu}-e^{i\mu}\right)\\
\mathcal{T}_{3(3,0)}^{-}\left(e^{3i\mu}-e^{-i\mu}\right)
\end{pmatrix}q^{3}+\begin{pmatrix}0\\
\mathcal{T}_{3(2,1)}^{-}\left(e^{i\mu}-e^{-i\mu}\right)
\end{pmatrix}q^{2}p\nonumber\\
 & +\begin{pmatrix}\mathcal{T}_{3(1,2)}^{+}\left(e^{-i\mu}-e^{i\mu}\right)\\
0
\end{pmatrix}qp^{2}+\begin{pmatrix}\mathcal{T}_{3(0,3)}^{+}\left(e^{-3i\mu}-e^{i\mu}\right)\\
\mathcal{T}_{3(0,3)}^{-}\left(e^{-3i\mu}-e^{-i\mu}\right)
\end{pmatrix}p^{3},
\end{align}
with $\mathcal{C}_{3(2,1)}^{+}=\mathcal{C}_{3(1,2)}^{-}=0$. This means that the terms $\mathcal{S}_{3,new(2,1)}^{+}$ and $\mathcal{S}_{3,new(1,2)}^{-}$ cannot be eliminated. All the other terms are eliminated by choosing 
\begin{equation}
\mathcal{T}_{3(k_{+},k_{-})}^{\pm}=\frac{-S_{3,new(k_{+},k_{-})}^{\pm}}{\left(e^{i\mu\left(k_{+}-k_{-}\right)}-e^{\pm i\mu}\right)}\quad\text{for}\quad k_{+}-k_{-}\ne\pm1.
\end{equation}

The values of the $\mathcal{T}_{3(k_{+},k_{-})}^{\pm}$ for the centrifugal governor example are given in table \ref{tab:T3}.

\begin{table}[!ht]
\centering
\caption{The values of the $\mathcal{T}_{3(k_{+},k_{-})}^{\pm}$. Note that $\mathcal{T}_{3(k_{+},k_{-})}^{+}=\mathcal{T}_{3(k_{-},k_{+})}^{-}$.}
\label{tab:T3}
\begin{tabular}{|c||c|r||c|r|}
\hline 
O & Coeff. & Value & Coeff. & Value\\
\hline 
\hline 
3 & $\mathcal{T}_{3(3,0)}^{+}$ & 0.102062073 &$\mathcal{T}_{3(3,0)}^{-}$& 0\\
\hline 
3 & $\mathcal{T}_{3(2,1)}^{+}$ & 0 &$\mathcal{T}_{3(2,1)}^{-}$& -0.306186218\\
\hline 
3 & $\mathcal{T}_{3(1,2)}^{+}$ &-0.306186218 & $\mathcal{T}_{3(1,2)}^{-}$& 0\\
\hline 
3 & $\mathcal{T}_{3(0,3)}^{+}$ & 0 &$\mathcal{T}_{3(0,3)}^{-}$&0.102062073\\
\hline 
\end{tabular}
\end{table}

After the third order transformation the resulting map is of the following form

\begin{align}
\mathcal{M}_{3} & =\underbrace{\begin{pmatrix}e^{i\mu} & 0\\
0 & e^{-i\mu}
\end{pmatrix}\begin{pmatrix}q_{3}\\
p_{3}
\end{pmatrix}}_{\mathcal{R}}+\underbrace{\begin{pmatrix}\mathcal{S}_{3,new(2,1)}^{+}\\
0
\end{pmatrix}q_{3}^{2}p_{3}+\begin{pmatrix}0\\
\mathcal{S}_{3,new(1,2)}^{-}
\end{pmatrix}q_{3}p_{3}^{2}}_{\mathcal{S}_{3,transformed}}\nonumber\\
 & =\begin{pmatrix}\left(e^{i\mu}+\mathcal{S}_{3,new(2,1)}^{+}q_{3}p_{3}\right)q_{3}\\
\left(e^{-i\mu}+\mathcal{S}_{3,new(1,2)}^{-}q_{3}p_{3}\right)p_{3}
\end{pmatrix}=\begin{pmatrix}f^{+}\left(q_{3}p_{3}\right)q_{3}\\
f^{-}\left(q_{3}p_{3}\right)p_{3}
\end{pmatrix}.\label{eq:M3}
\end{align}

The corresponding values for the coefficients can be found in table \ref{tab:M1} (linear) and table \ref{tab:M2} (third order). Note that the complex conjugate property $\mathcal{M}_{3}^{+}=\overline{\mathcal{M}}_{3}^{-}$ is maintained. 

Having calculated one (even) order without and one (uneven) order with surviving terms concludes the illustrative calculation of the order-by-order transformations for the example here, after the third order transformation. In principle the calculation of the transformations can be continued up to arbitrary order. With each transformation, the higher order terms are change and in the end only the terms $\mathcal{S}_{m(k,k-1)}^{+}$ and $\mathcal{S}_{m(k-1,k)}^{-}$ of uneven orders survive. Hence, the components $\mathcal{M}_{m}^{\pm}$ can also be factorize into the $f^{\pm}\left(q_{m}p_{m}\right)$ notation (eq. \ref{eq:M3}) for higher orders. 

\subsection{Transformation back to real space (normal form)\label{subsec:Normal-Form}}

Since the original map $\mathcal{M}_{0}$ only operates in real space, the normal form map $\mathcal{M}_\text{NF}$ should also only operate in real space. This is why the current map $\mathcal{M}_{m}$, where $m$ is the order of last transformation, is transformed to a real normal form basis $(q_\text{NF},p_\text{NF})$ composed of the real and imaginary parts of the current complex conjugate basis $\left(q_{m},p_{m}\right)$ \cite[eq 7.58+59+67]{Berz1999MMM}: 
\begin{eqnarray}
q_\text{NF} & = & \left(q_{m}+p_{m}\right)/\sqrt{2}\nonumber\\
p_\text{NF} & = & \left(q_{m}-p_{m}\right)/i\sqrt{2}\nonumber\\
q_{m} & = & \left(q_\text{NF}+i\,p_\text{NF}\right)/\sqrt{2}\nonumber\\
p_{m} & = & \left(q_\text{NF}-i\,p_\text{NF}\right)/\sqrt{2}\nonumber\\
q_{m}p_{m} & = & \frac{1}{2}\left(q_\text{NF}^{2}+p_\text{NF}^{2}\right)=\frac{1}{2}r_\text{NF}^{2}.\label{eq:qmpm}
\end{eqnarray}

The associated transfer matrix to the real normal form basis is obtained from the equations above 
\begin{equation}
\mathcal{A}_{real}=\frac{1}{\sqrt{2}}\begin{pmatrix}1 & 1\\
-i & i
\end{pmatrix}=\begin{pmatrix}\left(q_\text{NF}|q_{m}\right) & \left(q_\text{NF}|p_{m}\right)\\
\left(p_\text{NF}|q_{m}\right) & \left(p_\text{NF}|p_{m}\right)
\end{pmatrix}.
\end{equation}

The inverse relation is given accordingly
\begin{equation}
\mathcal{A}_{real}^{-1}=\frac{1}{\sqrt{2}}\begin{pmatrix}1 & i\\
1 & -i
\end{pmatrix}=\begin{pmatrix}\left(q_{m}|q_\text{NF}\right) & \left(q_{m}|p_\text{NF}\right)\\
\left(p_{m}|q_\text{NF}\right) & \left(p_{m}|p_\text{NF}\right)
\end{pmatrix}.
\end{equation}

The transformation back to the real space (into normal from space) yields
\begin{eqnarray}
\mathcal{M}_\text{NF} & = & \mathcal{A}_{real}\circ\mathcal{M}_{m}\circ\mathcal{A}_{real}^{-1}=\begin{pmatrix}\frac{1}{\sqrt{2}} & \frac{1}{\sqrt{2}}\\
\frac{-i}{\sqrt{2}} & \frac{i}{\sqrt{2}}
\end{pmatrix}\cdot\frac{1}{\sqrt{2}}\cdot\begin{pmatrix}f^{+}\left(r_\text{NF}^{2}\right)\left(q_\text{NF}+i\,p_\text{NF}\right)\\
f^{-}\left(r_\text{NF}^{2}\right)\left(q_\text{NF}-i\,p_\text{NF}\right)
\end{pmatrix}\nonumber\\
 & = & \begin{pmatrix}\frac{1}{2}\left(f^{+}+\bar{f}^{+}\right)t_{+}+\frac{i}{2}\left(f^{+}-\bar{f}^{+}\right)t_{-}\\
\frac{-i}{2}\left(f^{+}-\bar{f}^{+}\right)t_{+}+\frac{1}{2}\left(f^{+}+\bar{f}^{+}\right)t_{-}
\end{pmatrix}\nonumber\\
 & = & \begin{pmatrix}\text{Re}\left(f^{+}\left(r_\text{NF}^{2}\right)\right) & -\text{Im}\left(f^{+}\left(r_\text{NF}^{2}\right)\right)\\
\text{Im}\left(f^{+}\left(r_\text{NF}^{2}\right)\right) & \text{Re}\left(f^{+}\left(r_\text{NF}^{2}\right)\right)
\end{pmatrix}\cdot\begin{pmatrix}q_\text{NF}\\
p_\text{NF}
\end{pmatrix}.\label{eq:MNFReIm}
\end{eqnarray}

For the example of the centrifugal governor up to order three the normal form is
\begin{equation}
\mathcal{M}_\text{NF}=\begin{pmatrix}\cos\mu+\frac{1}{2}\text{Re}\left(\mathcal{S}_{3,new(2,1)}^{+}\right)r_\text{NF}^{2} & -\sin\mu-\frac{1}{2}\text{Im}\left(\mathcal{S}_{3,new(2,1)}^{+}\right)r_\text{NF}^{2}\\
\sin\mu+\frac{1}{2}\text{Im}\left(\mathcal{S}_{3,new(2,1)}^{+}\right)r_\text{NF}^{2} & \cos\mu+\frac{1}{2}\text{Re}\left(\mathcal{S}_{3,new(2,1)}^{+}\right)r_\text{NF}^{2}
\end{pmatrix}\cdot\begin{pmatrix}q_\text{NF}\\
p_\text{NF}
\end{pmatrix}.
\end{equation}

The table \ref{tab:MNF} below yields the values for the normal form map of our example case. 
\begin{table}[!ht]
\centering
\caption{The normal form map $\mathcal{M}_\text{NF}$ up to order three. The component $\mathcal{M}_\text{NF}^{+}$ is on the left, $\mathcal{M}_\text{NF}^{-}$ on the right.}
\label{tab:MNF}
\begin{tabular}{|c||c|r||c|r|}
\hline 
O & Coeff. & Value & Coeff. & Value\\
\hline \hline
1 & $\mathcal{M}_{NF(1,0)}^+$ & 0.339185989 & $\mathcal{M}_{NF(1,0)}^-$ & 0.940719334\\
\hline 
1 & $\mathcal{M}_{NF(0,1)}^+$ & -0.940719334 & $\mathcal{M}_{NF(0,1)}^-$ & 0.339185989\\
\hline \hline
3 & $\mathcal{M}_{NF(3,0)}^+$ & 0.470359667 & $\mathcal{M}_{NF(3,0)}^-$ & -0.169592994\\
\hline 
3 & $\mathcal{M}_{NF(2,1)}^+$ & 0.169592994 & $\mathcal{M}_{NF(2,1)}^-$ & 0.470359667\\
\hline 
3 & $\mathcal{M}_{NF(1,2)}^+$ & 0.470359667 & $\mathcal{M}_{NF(1,2)}^-$ & -0.169592994\\
\hline 
3 & $\mathcal{M}_{NF(0,3)}^+$ & 0.169592994 & $\mathcal{M}_{NF(0,3)}^-$ & 0.470359667\\
\hline 
\end{tabular}
\end{table}

The normal form transformation from $\mathcal{M}_{0}$ to $\mathcal{M}_\text{NF}$ can be obtained by the combination of all the single transformations yielding 
\begin{equation}
\mathcal{M}_\text{NF}=\underbrace{\mathcal{A}_{real}\circ\mathcal{A}_{m}\circ\mathcal{A}_{m-1}\circ...\circ\mathcal{A}_{1}\circ\mathcal{A}_{FP}}_{\mathcal{A}}\circ\mathcal{M}_{0}\circ\underbrace{\mathcal{A}_{FP}^{-1}\circ\mathcal{A}_{1}^{-1}\circ...\circ\mathcal{A}_{m-1}^{-1}\circ\mathcal{A}_{m}^{-1}\circ\mathcal{A}_{real}^{-1}}_{\mathcal{A}^{-1}}.\label{eq:NFTransf}
\end{equation}
The values of the coefficients of the full normal form transformation $\mathcal{A}$ are given in table \ref{tab:MNF_trans}.
\begin{table}[!ht]
\centering
\caption{The normal form transformation $\mathcal{A}$ up to order 3. The component $\mathcal{A}^{+}$ is on the left, $\mathcal{A}^{-}$ on the right.}
\label{tab:MNF_trans}
\begin{tabular}{|c||c|r||c|r|}
\hline 
O & Coeff. & Value & Coeff. & Value\\
\hline \hline
1 & $\mathcal{A}_{1(1,0)}^+$ & -1.106681920 & $\mathcal{A}_{1(1,0)}^-$ & 0\\
\hline 
1 & $\mathcal{A}_{1(0,1)}^+$ & 0 & $\mathcal{A}_{1(0,1)}^-$ & 0.903602004\\
\hline \hline
2 & $\mathcal{A}_{2(2,0)}^+$ & -0.319471552 & $\mathcal{A}_{2(2,0)}^-$ & 0\\
\hline 
2 & $\mathcal{A}_{2(1,1)}^+$ & 0 & $\mathcal{A}_{2(1,1)}^-$ & -0.521694860\\
\hline 
2 & $\mathcal{A}_{2(0,2)}^+$ & -0.425962069 & $\mathcal{A}_{2(0,2)}^-$ & 0\\
\hline \hline
3 & $\mathcal{A}_{3(3,0)}^+$ & 0.046111747 & $\mathcal{A}_{3(3,0)}^-$ & 0\\
\hline 
3 & $\mathcal{A}_{3(2,1)}^+$ & 0 & $\mathcal{A}_{3(2,1)}^-$ & 0.414150918\\
\hline 
3 & $\mathcal{A}_{3(1,2)}^+$ & 0.399635138  & $\mathcal{A}_{3(1,2)}^-$ & 0\\
\hline 
3 & $\mathcal{A}_{3(0,3)}^+$ & 0 & $\mathcal{A}_{3(0,3)}^-$ & -0.025100056\\
\hline 
\end{tabular}
\end{table}

Writing the complex conjugate functions $f^{\pm}$ from the equations above (particularly eq. \ref{eq:MNFReIm}) in a complex notation as $f^{\pm}\left(r_\text{NF}^{2}\right)=e^{\pm i\Lambda\left(r_\text{NF}^{2}\right)}$ illustrates circular behavior of the normal form:
\begin{equation}
\mathcal{M}_\text{NF}=\begin{pmatrix}\cos\left(\Lambda\left(r_\text{NF}^{2}\right)\right) & -\sin\left(\Lambda\left(r_\text{NF}^{2}\right)\right)\\
\sin\left(\Lambda\left(r_\text{NF}^{2}\right)\right) & \cos\left(\Lambda\left(r_\text{NF}^{2}\right)\right)
\end{pmatrix}\cdot\begin{pmatrix}q_\text{NF}\\
p_\text{NF}
\end{pmatrix}.
\end{equation}
It shows that the normal form $\mathcal{M}_\text{NF}$ consists of circular curves in phase space with only amplitude depended angle advancements $\Lambda\left(r_\text{NF}^{2}\right)$. 

\subsection{Angle advancement, tune and tune shifts}
In the beam physics terminology, the angle advancements $\Lambda\left(r_\text{NF}^{2}\right)$ are scaled to the interval $[0,1]$ instead of $[0,2\pi]$ and referred to as the tune and amplitude dependent tune shifts \cite{Berz1999MMM}.
The angle advancement can be calculated from the normal from map via
\begin{align}
\Lambda\left(r_\text{NF}^{2}=q_\text{NF}^{2},p_\text{NF}=0\right) & =\arccos\left(\frac{\left.\mathcal{M}_\text{NF}^{+}\right|_{p_\text{NF}=0}}{q_\text{NF}}\right)=\arccos\left(\text{Re}\left(f^{+}\left(r_\text{NF}^{2}\right)\right)\right).
\end{align}
For the centrifugal governor angle advancement is given by 
\begin{align}
\Lambda\left(r_\text{NF}^{2}=q_\text{NF}^{2}\right) & =\arccos\left(\cos\mu+\frac{1}{2}\text{Re}\left(\mathcal{S}_{3,new(2,1)}^{+}\right)r_\text{NF}^{2}\right)\nonumber\\
 & =\mu-\frac{\text{Re}\left(\mathcal{S}_{3,new(2,1)}^{+}\right)}{2\sin\mu}r_\text{NF}^{2}.
\end{align}

Note that $\mu$ is the eigenvalue phase of the original linear part. Accordingly, the tune is $\mu/2\pi$. For the centrifugal governor the tune and tune shifts are 
\begin{equation}
\frac{\Lambda\left(r_\text{NF}^{2}\right)}{2\pi}=0.1949242-0.07957747r_\text{NF}^{2}
\end{equation}

Furthermore, $r_\text{NF}^{2}$ can be expressed in terms of the original coordinates $(q_{0},p_{0})$
\begin{align}
r_\text{NF}^{2}\left(q_{0},p_{0}\right) & =\left(q_\text{NF}^{2}\left(q_{0},p_{0}\right)+p_\text{NF}^{2}\left(q_{0},p_{0}\right)\right)\nonumber\\
 & =\left(\mathcal{A}_{+}^{2}+\mathcal{A}_{-}^{2}\right)\left(q_{0},p_{0}\right).
\end{align}
Hence, 
\begin{equation}
\frac{\Lambda\left(q_{0},p_{0}\right)}{2\pi}=0.1949242-0.0974621q^{2}-0.0649747p^{2}-0.05626977q^{3}.
\end{equation}

\subsection*{Acknowledgments}

Many Thanks to Martin Berz for introducing me to the DA normal form algorithm, for our work, and our many discussions in this regard. This work was supported by the \textit{Studienstiftung des deutschen Volkes} with a scholarship to the author. 

\bibliographystyle{plain}
\bibliography{CG_Bib}

\begin{thebibliography}{1}

\bibitem{Berz1987PowerSeriesTracking}
Martin Berz.
\newblock The method of power series tracking for the mathematical description
  of beam dynamics.
\newblock {\em Nuclear Instruments and Methods A}, 258(3):431--436, 1987.

\bibitem{Berz1988DABeam}
Martin Berz.
\newblock Differential algebraic description of beam dynamics to very high
  orders.
\newblock {\em Part. Accel.}, 24(SSC-152):109--124, 1988.

\bibitem{Berz1999MMM}
Martin Berz.
\newblock {\em Modern Map Methods in Particle Beam Physics}.
\newblock Academic Press, 1999.

\bibitem{Courant1958TwissPara}
E.D Courant and H.S Snyder.
\newblock Theory of the alternating-gradient synchrotron.
\newblock {\em Annals of Physics}, 3(1):1 -- 48, 1958.

\bibitem{forest1989normal}
Etienne Forest, John Irwin, and Martin Berz.
\newblock Normal form methods for complicated periodic systems.
\newblock {\em Part. Accel.}, 24:91--107, 1989.

\bibitem{grote2006}
Johannes Grote, Martin Berz, and Kyoko Makino.
\newblock High-order representation of {P}oincaré maps.
\newblock {\em Nuclear Instruments and Methods A}, 558(1):106--111, 2006.

\end{thebibliography}

\section{Appendix\label{sec:Appendix}}
\subsection{Changes for higher dimensional symplectic phase space systems}
Considering an $n$ dimensional phase space system, the requirement for the fixed point property of the map and the calculation of the parameter dependent fixed point (eq. \ref{eq:PDFP1}+\ref{eq:PDFP2}) remain unchanged. We also still require that the map is linearly stable with distinct complex conjugate eigenvalue pairs. The diagonalization divides the map $\mathcal{M}_1(\vec{q}_1,\vec{p}_1)$ into $n$ subsystems identified by the distinct complex conjugate eigenvalue pairs
\begin{equation}
	\mathcal{M}_{1,j}\left(\vec{q}_1,\vec{p}_1\right)=
	\begin{pmatrix}\mathcal{M}_{1,j}^+\left(\vec{q}_1,\vec{p}_1\right)\\\mathcal{M}_{1,j}^-\left(\vec{q}_1,\vec{p}_1\right)
\end{pmatrix}
=\begin{pmatrix}e^{i\mu_j}+\sum_m\mathcal{S}_{m,j}^+\left({q}_{1,1},p_{1,1},{q}_{1,2},p_{1,2},...,{q}_{1,n},p_{1,n}\right)\\e^{-i\mu_j}+\sum_m \mathcal{S}_{m,j}^-\left({q}_{1,1},p_{1,1},{q}_{1,2},p_{1,2},...,{q}_{1,n},p_{1,n}\right)
\end{pmatrix}.\\
\end{equation} 
The polynomials $\mathcal{T}_m$ of the nonlinear transformations $\mathcal{A}_m$ are also considered in the $n$ subsystems separately, just like the commutator $\mathcal{C}_m=\left[\mathcal{T}_m,\mathcal{R}\right]$ with
\begin{eqnarray}
C_{m,j}&=&\sum_{||\vec{k}^{+}+\vec{k}^{-}||_{1}=m}\left(\mathcal{C}_{m,j}^{\pm}|\vec{k}^{+},\vec{k}^{-}\right)\prod_{l=1}^{n}q_{l}^{k_{l}^{+}}p_{l}^{k_{l}^{-}},\nonumber\\
\text{where}\quad\left(\mathcal{C}_{m,j}^{\pm}|\vec{k}^{+},\vec{k}^{-}\right)&=&\left(\mathcal{T}_{m,j}^{\pm}|\vec{k}^{+},\vec{k}^{-}\right)\left(e^{i\vec{\mu}\left(\vec{k}^{+}-\vec{k}^{-}\right)}-e^{\pm i\mu_{j}}\right)\nonumber\\
\text{and}\quad\left(\mathcal{T}_{m,j}^{\pm}|\vec{k}^{+},\vec{k}^{-}\right)&=&\frac{-\left(\mathcal{S}_{m,j}^{\pm}|\vec{k}^{+},\vec{k}^{-}\right)}{e^{i\vec{\mu}\left(\vec{k}^{+}-\vec{k}^{-}\right)}-e^{\pm i\mu_{j}}}.
\end{eqnarray}
The commutator has zero terms in subsystem $j$ and $(\mathcal{T}_{m,j}^{\pm}|\vec{k}^{+},\vec{k}^{-})$ not defined if \cite[eq. 7.65]{Berz1999MMM}
\begin{equation}
\text{mod}_{2\pi}\left(\mu_{j}(k_{j}^{+}-k_{j}^{-}\mp1)+\sum_{l\ne j}\mu_{l}\left(\vec{k}^{+}-\vec{k}^{-}\right)\right)=0,\label{eq:NF_ndim_condition}
\end{equation}
which means that the associated terms $(\mathcal{S}_{m,j}^{\pm}|\vec{k}^{+},\vec{k}^{-})$ cannot be eliminated and survive. A trivial solution of condition \ref{eq:NF_ndim_condition} is 
\begin{equation}
k_{j}^{+}-k_{j}^{-}	=	\pm1\quad\land\quad k_{l}^{+}=k_{l}^{-}\,\forall l\ne j\label{eq:trivialsurvival}
\end{equation}
where the first condition is for the (current) dimension $j$ (known from the one dimensional case) and the additional second condition is for the other dimensions $l$ with $l\ne j$. Note that the fundamental structure of the surviving terms from eq. \ref{eq:trivialsurvival} is the same as for the one dimensional case: After the $m$th order transformation the map has the following form (compare eq. \ref{eq:M3} and below)
\begin{equation}
\mathcal{M}_{m,j}\left(\vec{q}_m,\vec{p}_m\right)=
	\begin{pmatrix}\mathcal{M}_{m,j}^+\left(\vec{q}_m,\vec{p}_m\right)\\\mathcal{M}_{m,j}^-\left(\vec{q}_m,\vec{p}_m\right)
\end{pmatrix}=\begin{pmatrix}f_j^{+}\left(q_{m,1}p_{m,1},q_{m,2}p_{m,2},...,q_{m,n}p_{m,n}\right)q_{m,j}\\
f_j^{-}\left(q_{m,1}p_{m,1},q_{m,2}p_{m,2},...,q_{m,n}p_{m,n}\right)p_{m,j}\end{pmatrix},\label{eq:qmpmform}
\end{equation}
which is the rotationally invariant normal form with the quasi invariant normal form radius $r_\text{NF}$ (eq. \ref{eq:qmpm}) once transformed into real space following subsection \ref{subsec:Normal-Form}.

However, for $n>1$ the condition (eq. \ref{eq:NF_ndim_condition}) can also be satisfied due to resonances between the eigenvalue phases $\mu_i$ of the linear part. Terms that survive due to resonances do \textbf{not} fit into the structure of eq. \ref{eq:qmpmform} and therefore break the rotational symmetry of the resulting normal form. 

Consider $n=2$ with the order 7 resonance $2\mu_1-5\mu_2=-4\pi$, then the condition from eq. \ref{eq:NF_ndim_condition} is satisfied for $(k_1^+,k_1^-,k_2^+,k_2^-)=(1,0,0,5)$ and $(k_1^+,k_1^-,k_2^+,k_2^-)=(2,0,0,4)$, which means that the order six terms 
\begin{equation}
(\mathcal{S}_{6,1}^{-}|1,0,0,5)\quad\text{and}\quad(\mathcal{S}_{6,2}^{+}|2,0,0,4)
\end{equation}	
survive due to the resonance between $\mu_1$ and $\mu_2$ and break the rotational symmetry of the resulting normal form.

\subsection{Investigating the new third order terms produced by the second order transformation}
The following calculation investigates the term $\mathcal{S}_{2\rightarrow3}$
\begin{eqnarray}
\mathcal{S}_{2\rightarrow3} & =_{3} & \mathcal{S}_{2}\circ\left(\mathcal{I}-\mathcal{T}_{2}\right)-\mathcal{S}_{2}\nonumber \\
 & =_{3} & \mathcal{S}_{2(2,0)}\left(q-\mathcal{T}_{2}^{+}\right)^{2}+\mathcal{S}_{2(0,2)}\left(p-\mathcal{T}_{2}^{-}\right)^{2}\nonumber \\
 &  & +\left(\mathcal{S}_{2(1,1)}\left(q-\mathcal{T}_{2}^{+}\right)\left(p-\mathcal{T}_{2}^{-}\right)\right)_{3}-\mathcal{S}_{2}\nonumber \\
 & =_{3} & \mathcal{S}_{2(2,0)}q^{2}+\mathcal{S}_{2(1,1)}qp+\mathcal{S}_{2(0,2)}p^{2}-\mathcal{S}_{2}\nonumber \\
 &  & +\cancel{\mathcal{S}_{2(2,0)}\left(\mathcal{T}_{2}^{+}\right)^{2}+\mathcal{S}_{2(1,1)}\mathcal{T}_{2}^{+}\mathcal{T}_{2}^{-}+\mathcal{S}_{2(0,2)}\left(\mathcal{T}_{2}^{-}\right)^{2}}\nonumber \\
 &  & -2\mathcal{S}_{2(2,0)}\mathcal{T}_{2}^{+}q-\mathcal{S}_{2(1,1)}\left(\mathcal{T}_{2}^{+}p+\mathcal{T}_{2}^{-}q\right)-2\mathcal{S}_{2(0,2)}\mathcal{T}_{2}^{-}p\nonumber \\
 & =_{3} & -2\mathcal{S}_{2(2,0)}\mathcal{T}_{2}^{+}q-\mathcal{S}_{2(1,1)}\left(\mathcal{T}_{2}^{+}p+\mathcal{T}_{2}^{-}q\right)-2\mathcal{S}_{2(0,2)}\mathcal{T}_{2}^{-}p\label{eq:S23tot}
\end{eqnarray}
The surviving part of $\mathcal{S}_{2\rightarrow3}$ after the third order transformation is $\mathcal{S}_{2\rightarrow3\left(2,1\right)}^{+}$, which is complex conjugate to its counterpart $\mathcal{S}_{2\rightarrow3\left(1,2\right)}^{-}$
\begin{eqnarray}
\mathcal{S}_{2\rightarrow3\left(2,1\right)}^{+} & = & -2\mathcal{S}_{2(2,0)}^{+}\mathcal{T}_{2\left(1,1\right)}^{+}-\mathcal{S}_{2(1,1)}^{+}\left(\mathcal{T}_{2\left(2,0\right)}^{+}+\mathcal{T}_{2\left(1,1\right)}^{-}\right)-2\mathcal{S}_{2(0,2)}^{+}\mathcal{T}_{2\left(2,0\right)}^{-}\nonumber\\
 & = & \frac{2\mathcal{S}_{2(2,0)}^{+}\mathcal{S}_{2\left(1,1\right)}^{+}}{-e^{i\mu}}+\frac{\mathcal{S}_{2(1,1)}^{+}\mathcal{S}_{2\left(2,0\right)}^{+}}{e^{2i\mu}-e^{i\mu}}+\frac{\mathcal{S}_{2(1,1)}^{+}\mathcal{S}_{2\left(1,1\right)}^{-}}{-e^{-i\mu}}+\frac{\mathcal{S}_{2(0,2)}^{+}\mathcal{S}_{2\left(2,0\right)}^{-}}{e^{2i\mu}-e^{-i\mu}}
\end{eqnarray}

The calculation can be extended by calculating $\mathcal{S}_{2\rightarrow3\left(2,1\right)}^{+}$ in terms of $\mathcal{U}_{2}$ and the Twiss parameters by using the $\mathcal{S}_{2}=\mathcal{A}_{1}\circ\mathcal{U}_{2}\circ\mathcal{A}_{1}^{-1}$.

In the following calculation we are investigating the term $\mathcal{K}_{2\rightarrow3}$
\begin{align}
\mathcal{K}_{2\rightarrow3} & =\mathcal{T}_{2}\circ\left(\mathcal{R}-\mathcal{R}\circ\mathcal{T}_{2}+\mathcal{S}_{2}\right)-\mathcal{T}_{2}\circ\mathcal{R}\nonumber\\
 & =\mathcal{T}_{2}\circ\left(\mathcal{R}-\mathcal{K}_{2}\right)-\mathcal{T}_{2}\circ\mathcal{R}\nonumber\\
 & =_{3}\mathcal{T}_{2(2,0)}\left(e^{i\mu}q-\mathcal{K}_{2}^{+}\right)^{2}+\mathcal{S}_{2(0,2)}\left(e^{-i\mu}p-\mathcal{K}_{2}^{-}\right)^{2}\nonumber\\
 & \quad+\left(\mathcal{S}_{2(1,1)}\left(e^{i\mu}q-\mathcal{K}_{2}^{+}\right)\left(e^{-i\mu}p-\mathcal{K}_{2}^{-}\right)\right)_{3}-\mathcal{T}_{2}\circ\mathcal{R}\nonumber\\
 & =_{3}\mathcal{T}_{2(2,0)}e^{2i\mu}q^{2}+\mathcal{T}_{2(1,1)}qp+\mathcal{T}_{2(0,2)}e^{-2i\mu}p^{2}-\mathcal{T}_{2}\circ\mathcal{R}\nonumber\\
 & \quad+\cancel{\mathcal{T}_{2(2,0)}\left(\mathcal{K}_{2}^{+}\right)^{2}+\mathcal{T}_{2(1,1)}\mathcal{K}_{2}^{+}\mathcal{K}_{2}^{-}+\mathcal{T}_{2(0,2)}\left(\mathcal{K}_{2}^{-}\right)^{2}}\nonumber\\
 & \quad-2\mathcal{T}_{2(2,0)}\mathcal{K}_{2}^{+}e^{i\mu}q-\mathcal{T}_{2(1,1)}\left(\mathcal{K}_{2}^{+}e^{-i\mu}p+\mathcal{K}_{2}^{-}e^{i\mu}q\right)-2\mathcal{T}_{2(0,2)}\mathcal{K}_{2}^{-}e^{-i\mu}p
\end{align}
where
\begin{equation}
\mathcal{K}_{2} =\mathcal{R}\circ\mathcal{T}_{2}-\mathcal{S}_{2}\qquad\rightarrow\qquad\mathcal{K}_{2}^{\pm}=e^{\pm i\mu}\mathcal{T}_{2}^{\pm}-\mathcal{S}_{2}^{\pm}
\end{equation} 
so
\begin{align}
\mathcal{K}_{2\rightarrow3} & =_{3}2\mathcal{T}_{2(2,0)}\mathcal{S}_{2}^{+}e^{i\mu}q+\mathcal{T}_{2(1,1)}\left(\mathcal{S}_{2}^{+}e^{-i\mu}p+\mathcal{S}_{2}^{-}e^{i\mu}q\right)+2\mathcal{T}_{2(0,2)}\mathcal{S}_{2}^{-}e^{-i\mu}p\nonumber\\
 & \quad-2\mathcal{T}_{2(2,0)}\mathcal{T}_{2}^{+}e^{2i\mu}q-\mathcal{T}_{2(1,1)}\left(\mathcal{T}_{2}^{+}p+\mathcal{T}_{2}^{-}q\right)-2\mathcal{T}_{2(0,2)}\mathcal{T}_{2}^{-}e^{-2i\mu}p
 \end{align}
 The surviving part of $\mathcal{S}_{2\rightarrow3}$ after the third order transformation is $\mathcal{K}_{2\rightarrow3\left(2,1\right)}^{+}$, which is complex conjugate to its counterpart $\mathcal{K}_{2\rightarrow3\left(1,2\right)}^{-}$
 \begin{align}
\mathcal{K}_{2\rightarrow3\left(2,1\right)}^{+} & =2\mathcal{T}_{2(2,0)}^{+}\mathcal{S}_{2\left(1,1\right)}^{+}e^{i\mu}+\mathcal{T}_{2(1,1)}^{+}\left(\mathcal{S}_{2\left(2,0\right)}^{+}e^{-i\mu}+\mathcal{S}_{2\left(1,1\right)}^{-}e^{i\mu}\right)+2\mathcal{T}_{2(0,2)}^{+}\mathcal{S}_{2\left(2,0\right)}^{-}e^{-i\mu}\nonumber\\
 & \quad-2\mathcal{T}_{2(2,0)}^{+}\mathcal{T}_{2\left(1,1\right)}^{+}e^{2i\mu}-\mathcal{T}_{2(1,1)}^{+}\left(\mathcal{T}_{2\left(2,0\right)}^{+}+\mathcal{T}_{2\left(1,1\right)}^{-}\right)-2\mathcal{T}_{2(0,2)}^{+}\mathcal{T}_{2\left(2,0\right)}^{-}e^{-2i\mu}\nonumber\\
 & =\frac{-2\mathcal{S}_{2(2,0)}^{+}\mathcal{S}_{2\left(1,1\right)}^{+}}{e^{i\mu}-1}+\frac{\mathcal{S}_{2(1,1)}^{+}}{e^{i\mu}}\left(\mathcal{S}_{2\left(2,0\right)}^{+}e^{-i\mu}+\mathcal{S}_{2\left(1,1\right)}^{-}e^{i\mu}\right)+\frac{2\mathcal{S}_{2(0,2)}^{+}\mathcal{S}_{2\left(2,0\right)}^{-}}{e^{2i\mu}-e^{-i\mu}}
\end{align}

The calculation can be extended by expressing $\mathcal{T}_{2}^{\pm}$ in terms of $\mathcal{S}_{2}^{\pm}$ (see eq. \ref{eq:T2(S2)}) and further by expressing $\mathcal{S}_{2}$ in terms of $\mathcal{U}_{2}$ and the Twiss parameters by using $\mathcal{S}_{2}=\mathcal{A}_{1}\circ\mathcal{U}_{2}\circ\mathcal{A}_{1}^{-1}$ as above.
\end{document}